\documentstyle[12pt]{article}
\newcommand{\be}{\begin{eqnarray}}
\newcommand{\ee}{\end{eqnarray}}
\newcommand{\bdm}{\begin{displaymath}}
\newcommand{\edm}{\end{displaymath}}

\begin{document}

\begin{center}
{{\bf \LARGE{Localization of gravity on a de Sitter thick braneworld\\
\vskip 3mm
without scalar fields}}}
\end{center}

\vskip 1mm

\author{autor}
\begin{center}
{\bf \large {Alfredo Herrera--Aguilar\footnote{E-mail:
herrera@ifm.umich.mx}}}, \ {\bf \large {Dagoberto
Malag\'on--Morej\'on\footnote{E-mail: malagon@ifm.umich.mx}}},
\end{center}
\begin{center}
{\bf \large {and \ Refugio Rigel Mora--Luna\footnote{E-mail:
topc@ifm.umich.mx}}}
\end{center}


\begin{center}
Instituto de F\'{\i}sica y Matem\'{a}ticas, Universidad
Michoacana de San Nicol\'as de Hidalgo. \\
Edificio C--3, Ciudad Universitaria, C.P. 58040, Morelia,
Michoac\'{a}n, M\'{e}xico.
\end{center}

\begin{abstract}
In this work we present a simple thick braneworld model that is
generated by an intriguing interplay between a 5D cosmological
constant with a de Sitter metric induced in the 3--brane without the
inclusion of scalar fields. We show that 4D gravity is localized on
this brane, provide analytic expressions for the massive
Kaluza--Klein (KK) fluctuation modes and also show that the spectrum
of metric excitations displays a mass gap. We finally present the
corrections to Newton's law due to these massive modes. This model
has no naked singularities along the fifth dimension despite the
existence of a mass gap in the graviton spectrum as it happens in
thick branes with 4D Poincar\'e symmetry, providing a simple model
with very good features: the curvature is completely smooth along
the fifth dimension, it localizes 4D gravity and the spectrum of
gravity fluctuations presents a mass gap, a fact that rules out the
existence of phenomenologically dangerous ultralight KK excitations
in the model. We finally present our solution as a limit of scalar
thick branes.
\end{abstract}

\section*{The model and its physical consistency}

Within  the framework  of  the braneworld models embedded in a
spacetime with extra dimensions and after the success of the thin
brane models, where singularities are present at the position of the
branes \cite{akama}--\cite{Localgravity1}, in solving the mass
hierarchy and 4D gravity localization problems, it has became a
matter of interest to find smooth braneworld solutions. In some
models, such solutions are obtained by introducing one or several
scalar fields in the bulk \cite{dewolfe}--\cite{ds} and the large
variety of scalar fields that can be used to generate these models
gives rise to different scenarios without singularities at the
position of the branes. By following the line of smoothing out the
brane configurations we show that 4D gravity can be localized on a
de Sitter $3$--brane generated in five dimensions by gravity with a
positive cosmological constant only, avoiding at all the use of
scalar fields. Thus, our thick brane is not made of scalar matter
but rather is modeled by an intriguing relation between the
curvatures generated by the 5D and 4D cosmological constants.

In this scenario, the inclusion of the 5D cosmological constant
renders a completely smooth 5D manifold in the presence of a mass
gap in the KK graviton spectrum of metric fluctuations, in contrast
with previous results found in the context of thick braneworlds with
4D Poincar\'e symmetry \cite{gremm}, \cite{hmmn}.

The complete action for the braneworld model is expressed as follows
\begin{equation}
S = 2M^3\int d^5 x \sqrt{-g} \left( R -2\Lambda_5\right).
\label{accion}
\end{equation}
It describes 5D gravity with a bulk cosmological constant
$\Lambda_5,$ and $M$ is the gravitational coupling constant in 5D
(in fact we can write $\frac{1}{2\kappa_5^2}$ instead of $2M^3$).

The Einstein equations with a cosmological constant in five
dimensions are given by
\begin{equation}
 G_{AB} =-\Lambda_5g_{AB}
\label{einequ}
\end{equation}
For the background metric we use the ansatz of a warped 5D line
element with an induced $3$--brane with spatially flat cosmological
background that reads
\begin{equation}
ds^2 = e^{2f(\sigma)} \left[- d t^2 + a^2(t) \left(d x^2 + d y^2 + d
z^2 \right)  \right] + d \sigma^2 \label{ansatz}
\end{equation}
where $f(\sigma)$ is the warp factor and $a(t)$ is the scale factor
of the brane.

By using the ansatz (\ref{ansatz}) we can compute the components of
the Einstein tensor
\begin{eqnarray}
G_{00} &=&   3 \left[\frac{\dot a^2}{a^2} - e^{2 f}\left( 2f^{'2} +
f^{''}\right)\right],
\label{eqeintach1}\nonumber \\
G_{\alpha\alpha} &=& - 2\ddot aa - \dot a^2 + 3a^2e^{2 f} \left( 2f^{'2} + f^{''} \right) \label{eqeintach2},\nonumber\\
G_{\sigma\sigma} &=& -3e^{-2 f} \left(\frac{\ddot a}{a} + \frac{\dot
a^2}{a^2} \right) + 6f^{'2}, \label{eqeintach3}
\label{einteintensor}
\end{eqnarray}
where ${"\prime"}$ and ${"\cdotp"}$ are derivatives with respect to
the extra dimension and time, respectively.

The set of equations (\ref{einequ}) can be rewritten in a simple
way:
\begin{eqnarray}
f^{''} &=&\frac{1}{3}\left(2\frac{\dot a^2}{a^2}-5\frac{\ddot a}{a}
\right)e^{-2 f}, \label{einsteinequ}
\\f^{'2} &=&\frac{1}{6}\left[\left(5\frac{\ddot a}{a}+\frac{\dot
a^2}{a^2} \right)-\Lambda_5\right] e^{-2 f}. \label{restriccion}
\end{eqnarray}

If we derive the restriction equation (\ref{restriccion}) with
respect to the extra dimension in order to compare it to
(\ref{einsteinequ}), we get a second order differential equation for
the scale factor with a general solution of the following type
$$a(t)=ce^{Ht},$$
where $c$ and $H$ are arbitrary constants. Thus, mathematical
consistency of the Einstein equations forces us to have a scale
factor corresponding to a de Sitter 4D cosmological background
defined by $a(t)=e^{Ht},$ since the constant $c$ can be absorbed
into a coordinate redefinition. Once we have determined the form of
the scale factor, the solution for the warp factor is
straightforward and reads
\begin{eqnarray}
f(\sigma)=\ln\left(\frac{H}{b}\cos(b\sigma)\right), \label{soln}
\end{eqnarray}
where $-\pi/2\le\sigma\le\pi/2,$ the constant $b$ is inversely
proportional to the thickness of the $3$--brane and is determined by
the 5D cosmological constant as follows:
\begin{equation}
b^2=\frac{\Lambda_5}{6}. \label{bb2}
\end{equation}

By performing the following coordinate transformation
\begin{equation}
z=\int e^{-f(\sigma)}d\sigma, \label{ct}
\end{equation}
the 5D metric (\ref{ansatz}) gets an overall conformal warp factor
\begin{equation}
 ds^{2}=e^{2f(z)}\left[g_{\mu \nu}dx^{\mu}dx^{ \nu}+dz^{2}\right],
\end{equation}
or, in terms of the metric $g_{MN}=e^{2f} \bar g_{MN},$ which
enables us to easily compute the 5D curvature scalar $R$ in terms of
the $\bar R$ and the warp factor through the known formula
\cite{wald}
\begin{equation}
R=e^{-2f}\left[ \bar R -2(d-1)\bar g^{MN} \bar \bigtriangledown_{M}
\bar \bigtriangledown_{N} f - (d-2)(d-1)g^{MN} \bar
\bigtriangledown_{M} f \bar \bigtriangledown_{N} f \right],
\end{equation}
where $d$ is the total number of dimensions. This fact, in turn,
allows us to separate an effective 4D action from the 5D one
\begin{equation}
S_{eff}\supset \int d^{4}x\sqrt{-\bar g}\left\{
2M^{3}\int^{\infty}_{-\infty}\left[e^{3f}\bar R +
4H^{2}e^{3f}\left(5\,{\rm sech}^{2}(Hz)-3\right)-2e^{5f}\Lambda_5
\right]dz\right\}. \label{Seff}
\end{equation}
After integrating over the extra coordinate $z(\sigma)$ we shall
make a direct comparison of the result to the 4D Einstein--Hilbert
action on the brane
\begin{equation}
S_{brane} = 2M^2_{pl}\int d^4 x \sqrt{-^{4}g} \label{4Daction}
\left(^{4}R-2\Lambda_4\right).
\end{equation}

Thus, in order to derive the scale of 4D gravitational interactions
we look at the curvature term and perform the following integration:
\begin{equation}
M^2_{Pl}=M^3\int^{\infty}_{-\infty} e^{3f(z)}dz=\frac{M^3H^3}{b^3}
\int^{\infty}_{-\infty} \,{\rm sech}^3(Hz)dz=\frac{\pi
M^3H^2}{2b^3}=\frac{3\sqrt{6}\pi M^3H^2}{\Lambda_5^{3/2}},
\end{equation}
which turns out to be finite, as it should be for a well defined 4D
theory.

The second and third terms in the integral (\ref{Seff}) contribute
to the definition of the 4D cosmological constant $\Lambda_4$
\begin{equation}
4M^2_{pl}\Lambda_{4}=\int^{\infty}_{-\infty}
\left[8M^{3}H^{2}e^{3f}\left(3-5{\rm sech}^{2}(Hz)\right)+
4M^{3}e^{5f}\Lambda_5\right]dz.
\end{equation}
After integrating along the fifth dimension and simplifying the
resulting expression we get
\begin{equation}
 \Lambda_{4}=3H^{2}.
\end{equation}
This relation can also be obtained from the Friedmann equations
derived by directly varying the action (\ref{4Daction}).

\section*{Metric perturbations, localization of 4D gravity and
corrections to Newton's law} In order to compute the corrections to
Newton's law we need to perform some preliminary work. First of all
we perturb the ansatz for the metric and then we compute the
Einstein's equations to first order. For the sake of simplifying the
calculations we use the freedom we have to impose an axial gauge
$h_{5M}=0$ on the perturbed metric:
\begin{equation}
ds^{2}=e^{2f(\sigma)}\left[g_{\mu \nu}+h_{\mu \nu}
\right]dx^{\mu}dx^{\nu}+d\sigma^{2}. \label{metricz}
\end{equation}

After imposing the transverse traceless condition
$h^{\mu}_{\mu}=\partial^{\mu}h_{\mu\nu}=0$, the system of perturbed
Einstein equations to first order adopts the form:
\begin{equation}
\left[\partial^{2}_{\sigma}+4f^{'}\partial_{\sigma}+e^{-2f(\sigma)}
\left(-\partial^{2}_{t}-3H\partial_{t}+e^{-2Ht}\nabla^2-2H^2 \right)
\right]\bar h_{\mu \nu}=0, \label{per1}
\end{equation}
where $\nabla^2$ is the Laplacian in flat space and $\bar h_{\mu
\nu}$ are the transverse traceless modes of the metric fluctuations.

In order to simplify calculations we perform the coordinate
transformation (\ref{ct}) which allows us to study in detail the
physics of the previous equation. Thus, the equation (\ref{per1})
takes the following form
\begin{equation}
\left[\partial^{2}_{z}+3f^{'}\partial_{z}-\partial^{2}_{t}-3H\partial_{t}+e^{-2Ht}\nabla^2
-2H^2\right]\bar h_{\mu \nu}=0. \label{ecusep}
\end{equation}
At this point we shall make use of the following separation of
variables
\begin{equation}
\bar h_{\mu \nu}=e^{-\frac{3}{2}f(z)}\Psi_{\mu\nu}(z)g(x).
\label{sepvar}
\end{equation}

This change of variables leads to a Schr\"odinger--like equation
(where the indices of the function $\Psi_{\mu\nu}(z)$ have been
omitted for convenience) for the 5D sector:
\begin{equation}
\left(-\partial^{2}_{z}+\frac{9}{4}f^{'2}+\frac{3}{2}f^{''}-m^{2}
\right)\Psi(z)=0,
\end{equation}
where one can identify the quantum mechanical potential $V_{QM}$ as
follows
\begin{equation}
V_{QM}=\frac{9}{4}f^{'2}+\frac{3}{2}f^{''}.
\end{equation}

Finally, the 4D equation that we obtain reads
\begin{equation}
\left(-\partial^{2}_{t}-3H\partial_{t}+e^{-2Ht}\nabla^2-2H^2
\right)g(x)=-m^{2}g(x),
\end{equation}
where the integration constant $m^{2}$ is defined as the mass in a
4D de Sitter spacetime \cite{dsmass}.

The warp factor in the language of the variable $z$ is given by
\begin{equation}
f={\rm ln}\left(\frac{H}{b}{\rm sech}(Hz)\right).
\end{equation}

Thus, in terms of this coordinate, our quantum mechanical potential
takes the form of a modified P\"oschl--Teller one
\begin{equation}
V_{QM}=-\frac{15}{4}H^{2}{\rm sech}^{2}(Hz)+\frac{9}{4}H^{2},
\label{VQM}
\end{equation}
which ensures the existence of a mass gap determined by its
asymptotic value, namely, by $\frac{9}{4}H^{2},$ or equivalently
$m=\frac{3H}{2}$.

Thus, we have a classical eigenvalue problem for the following
Schr\"odinger equation
\begin{equation}
 \left(-\partial^{2}_{z}-\frac{15}{4}H^{2}{\rm sech}^{2}(Hz)+\frac{9}{4}H^{2}-m^{2} \right)\Psi(z)=0,
\end{equation}
with the general solution given by
\begin{equation}
 \Psi(z)=C_1\,P^{\mu}_{\frac{3}{2}}\left(\tanh(Hz)\right)+
 C_2\,Q^{\mu}_{\frac{3}{2}}\left(\tanh(Hz)\right)
\end{equation}
where $P^{\mu}_{\frac{3}{2}}$ and $Q^{\mu}_{\frac{3}{2}}$ are
associated Legendre functions of first and second kind,
respectively, of degree $\nu=3/2$ and order
$\mu=\sqrt{\frac{9}{4}-\frac{m^{2}}{H^{2}}}$ \cite{bhnqrs}.

Thus, the last differential equation has two discrete bound states:
the first one corresponding to a massless bound state with energy
$E_0=-\frac{9}{4}H^{2}$ and $\mu=3/2$
\begin{equation}
 \Psi_{0}(z)=k_{0}\,{\rm sech}^{\frac{3}{2}}(Hz),
\end{equation}
physically interpreted as a stable graviton localized on the brane,
and a second one corresponding to an excited state with
$m^{2}=2H^{2},$ energy $E_{1}=-\frac{1}{4}H^2$ and $\mu=\frac{1}{2}$
\begin{equation}
 \Psi_{1}(z)=k_{1}\sinh(Hz)\,{\rm sech}^{\frac{3}{2}}(Hz),
\end{equation}
which represents a massive graviton also localized on the brane;
finally we have a tower of massive continuous modes that start from
$m\geq\frac{3H}{2}$ described by the following eigenfunctions with
imaginary order $\mu=i\rho$ \cite{bhnqrs}
\begin{equation}
\Psi_{m}(z)=C_{1}\,P^{\pm i\rho}_{\frac{3}{2}}\left(\tanh(Hz)
\right)+C_{2}\,Q^{\pm i\rho}_{\frac{3}{2}}\left(\tanh(Hz) \right),
\label{massmod}
\end{equation}
where $\rho=\sqrt{\frac{m^{2}}{H^{2}}-\frac{9}{4}},$ that must
asymptote to plane waves.

This result prepares to us the way to compute corrections to
Newton's law, since the massive modes contribute to Newton's law
with small corrections coming from the fifth dimension. In order to
find these corrections we consider $2m > 3H$ and $C_{2}=0$ in
equation (\ref{massmod}) since the continuous spectrum of
eigenfunctions asymptotically describes plane waves
\begin{equation}
\Psi_{\pm}^{\mu}(z)=C_{\pm}P^{\pm i\rho}_{\frac{3}{2}}
\left(\tanh(Hz) \right)\sim  \frac{1}{\sqrt{2\pi}}e^{\pm iH\rho z}.
\end{equation}
After this, we further take the thin brane limit $H\longmapsto
\infty$ and locate a probe mass in the center of the brane in the
transverse direction. The  corrections to the potential generated by
massive gravitons can be expressed as follows \cite{csakietal}
\begin{eqnarray}
U(r)\sim
\frac{M_{1}M_{2}}{r}\left(G_{4}+M_{*}^{-3}e^{-m_{1}r}
\left|\Psi_{1}(z_{0})\right|^{2}+M_{*}^{-3}\int_{m_{0}}^{\infty}dme^{-mr}
\left|\Psi^{\mu(m)}(z_{0})\right|^{2}\right)
\\=\frac{M_{1}M_{2}}{r}\left(G_{4}+\triangle G_{4}\right),
\end{eqnarray}
where the brane is located at $z = z_{0}$, $G_{4}$ is the
gravitational coupling constant in 4D, $\Psi_{1}(z_{0})$ is the wave
function of the first excited state and since it is an odd function,
it does not contribute in the thin brane limit that we are
considering, finally, $\Psi^{\mu}(z_{0})$ represent the continuous
massive states that need to be integrated over their masses. Thus,
the corrections to Newton's law contributed by the massive
continuous states have the following form  (see \cite{bhnqrs} for
details):
\begin{equation}
\triangle G_{4}\sim M_{*}^{-3}\frac{1}{\left|
\Gamma(-\frac{1}{4})\Gamma(\frac{7}{4})
\right|^2}\frac{e^{-\frac{3}{2}Hr}}{r}\left(1+O\left(\frac{1}{Hr}\right)\right),
\end{equation}
indicating its small character.

It is worth mentioning that in our setup there also exists a
relation between the ground bound state
$\Psi_{0}(z)=k_{0}e^{\frac{3}{2}f(z)},$ which ensures 4D gravity
localization, the smooth character of the 5D Ricci scalar $R$ and
the behaviour of the quantum mechanical potential $V_{QM}$ in the
same spirit as it was done for flat thick 3--branes in \cite{hmmn},
where it was shown that if one requires the existence of a mass gap
in the KK graviton spectrum along with the localization of 4D
gravity, then the 5D manifold necessarily develops naked
singularities at its boundaries. Conversely, if one imposes a smooth
character of the Ricci scalar together with 4D gravity localization,
then the quantum mechanical potential asymptotically vanishes, a
fact that rules out the presence of a mass gap in the graviton
spectrum of KK excitations. In this work we include a positive bulk
cosmological constant $\Lambda_5$ together with an induced flat FRW
metric (of de Sitter type, in particular) on the 3--brane.

It is easy to show that for the setup (\ref{accion}) with the ansatz
(\ref{metricz}) and for the potential $V_{QM}$ considered in
(\ref{VQM}), for smooth configurations at the position of the branes
we can write the following relation
\begin{equation}
R\Psi^{\frac{4}{3}}_{0} \sim \left[3\left(\frac{\ddot
a}{a}+\frac{\dot a^{2}}{a^{2}}\right)-\frac{8}{3}V_{QM}\right]
\end{equation}
which apart from the quantities included in \cite{hmmn}, it involves
the scale factor. Thus, from this expression we see that if one
requires the presence of a mass gap in the graviton spectrum of KK
excitations, the potential $V_{QM}$ must asymptotically adopt a
positive value; if indeed one requires 4D gravity to be localized,
which means that $\Psi_{0}\longrightarrow 0$ as
$z\longrightarrow\infty,$ then the Ricci scalar can be either finite
or singular depending on the asymptotic behaviour of the expression
in brackets: if it asymptotically vanishes, then the Ricci scalar
can be finite along the fifth dimension, while if it asymptotically
tends to a finite or infinite value, then the curvature scalar will
possess naked singularities at the boundaries of the manifold. A
similar result was recently obtained in \cite{glwf} for AdS$_5$ bulk
geometries with both dS and AdS induced metrics on the 3--brane,
where some examples are displayed as clarifying illustrations.

In our case the scale factor has the particular form corresponding
to a de Sitter metric $a=e^{Ht},$ yielding
\begin{equation}
R\Psi^{\frac{4}{3}}_{0} \sim \left[3H^2-\frac{4}{3}V_{QM}\right];
\end{equation}
since the quantum mechanical potential $V_{QM}$ asymptotically
approaches the positive value $9H^2/4$, ensuring the presence of a
mass gap in the graviton spectrum of KK fluctuations, then the
expression in brackets asymptotically vanishes, implying that the
scalar curvature is regular along the fifth dimension, which is
indeed the case since it is constant $R=10\Lambda_5/3$.

Thus, this result guarantees the smoothness of our braneworld
obtained without the use of scalar fields even in the presence of a
mass gap in the spectrum of KK graviton fluctuations, an important
phenomenological fact that rules out the existence of the dangerous
arbitrarily light KK excitations.

\section*{The solution as a limit of scalar thick branes}

The above presented solution (\ref{soln}) can also be obtained as a
special case or limit of scalar field thick branes, with a bent de
Sitter 4d metric, by fixing one parameter of these solutions.

By considering the action of a bulk scalar field $\phi$ with a
self--interacting potential $V(\phi)$ minimally coupled to 5d
gravity
\begin{equation}
S_5=\int
d^5x\sqrt{|G|}\left[\frac{1}{4}R_5-\frac{1}{2}(\nabla\phi)^2-V(\phi)\right],
\label{action}
\end{equation} and assuming a de Sitter $3$--brane in the metric ansatz
(\ref{ansatz}), one gets the following solution reported by Gremm
\cite{gremm} (see also \cite{dewolfe},\cite{wang}--\cite{kks})
\begin{eqnarray}
\label{sol1} e^{2A}=\cos^2\left(c\sigma\right),\qquad \qquad
\phi=\frac{1}{c}\sqrt{\frac{3}{2}\left(c^2-H^2\right)}\
\ln\left[\frac{1+\tan\left(\frac{c\sigma}{2}\right)}
{1-\tan\left(\frac{c\sigma}{2}\right)}\right],\\
V(\phi)=\frac{3}{4}
\cosh^2\left(\frac{c\phi}{\sqrt{\frac{3}{2}\left(c^2-H^2\right)}}\right)
\left[3H^2+c^2-4c^2\tanh^2{\left(\frac{c\phi}{\sqrt{\frac{3}{2}
\left(c^2-H^2\right)}}\right)}\right],\nonumber
\end{eqnarray}
where $c$ is an arbitrary constant. The 5d manifold with this choice
of $A(\sigma)$ has naked singularities at $\sigma=\pm\pi/(2c)$ and
the scalar field diverges at this singular point. However, in the
particular case $c=H,$ the scalar field vanishes and the potential
becomes constant, yielding the action (\ref{accion}) with the
corresponding solution (\ref{soln}).

It should be mentioned that when the author of \cite{gremm} analyses
the dynamics of metric fluctuations, he does not properly define the
mass in a 4d de Sitter spacetime and, consequently, the
corresponding Schr\"odinger equation possesses an analog quantum
mechanical potential which vanishes asymptotically when $c=H,$
however, he correctly reproduces the pair of bound states present in
the model for this limit.

On the other hand, Wang \cite{wang} has obtained a family of
solutions for the same field system (\ref{action}), which was
borrowed from a 4d solution presented in \cite{goetz}, and in terms
of the $z$ coordinate reads:
\begin{eqnarray}
\label{sol2} A=-n\ln\left[\cosh\left(cz\right)\right],\qquad
\phi=\phi_0\sin^{-1}\left[\tanh\left(cz\right)\right],\qquad
V(\phi)=V_0\cos^{2(1-n)}\left(\frac{\phi}{\phi_0}\right),
\end{eqnarray}
where $\phi_0=\sqrt{3n(1-n)},$ $V_0=nc^2[3(1+3n)]/2,$ $c$ and $n$
are arbitrary constants and the latter is subject to $0<n<1$.

Because of this inequality, in principle, this family of solutions
does not contain our particular solution as a special case, but as a
limit $n\rightarrow 1,$ when the scalar field vanishes and the
potential becomes constant.

The 5d curvature scalar for this solution reads
\begin{equation}
R=4c^2n(3n+2)\cosh^{2(n-1)}{cz}
\end{equation}
which is regular for $n\le 1$ and diverges for $n>1$; it is worth
noticing that it becomes constant in the limit $n\rightarrow 1,$ a
fact which agrees with the analysis performed in the previous
section regarding the singular or regular character of the 5d
manifold.

Finally, it is worth noticing that corrections to Newton's law were
not computed either in \cite{gremm} nor in \cite{wang}.

\section*{Acknowledgements} The authors acknowledge fruitful and
illuminating discussions with U. Nucamendi. This research was
supported by grants CIC--4.16 and CONACYT 60060--J. DMM and RRML
acknowledge PhD grants from CONACYT and UMSNH. AHA thanks SNI for
support.


\end{document}